\documentclass{aip-cp}

\usepackage[numbers]{natbib}
\usepackage{rotating}
\usepackage{graphicx}

\begin{document}

\title{Techniques And Results For The Calibration Of The MST Prototype For The Cherenkov Telescope Array}

\author[aff1]{L. Oakes\corref{cor1}}
\author[aff2]{M. Garczarczyk}
\author[aff1]{S. Kaphle}
\author[aff1]{M. Mayer}
\author[aff2]{S. Schlenstedt}
\author[aff1]{U. Schwanke}
\author[aff3]{the CTA Consortium}

\affil[aff1]{Humboldt-Universit\"at zu Berlin}
\affil[aff2]{DESY Zeuthen}
\affil[aff3]{Full consortium author list at www.cta-observatory.org}
\corresp[cor1]{loakes@physik.hu-berlin.de}

\maketitle

\begin{abstract}
The next generation instrument for ground-based gamma-ray astronomy will be the Cherenkov Telescope Array (CTA), consisting of approximately 100 telescopes in three sizes, built on two sites with one each in the Northern and Southern Hemispheres. Up to 40 of these will be Medium Size Telescopes (MSTs) which will dominate sensitivity in the core energy range. Since 2012, a full size mechanical prototype for the modified 12 m Davies-Cotton design MST has been in operation in Berlin. This document describes the techniques which have been implemented to calibrate and optimise the mechanical and optical performance of the prototype, and gives the results of over three years of observations and measurements. Pointing calibration techniques will be discussed, along with the development of a bending model,  and calibration of the CCD cameras used for pointing measurements.  Additionally alignment of mirror segments using the Bokeh method is shown.
\end{abstract}

\section{INTRODUCTION}
The next generation observatory for ground based gamma-ray astronomy will be the Cherenkov Telescope Array. Current Imaging Atmospheric Cherenkov Telescope (IACT) observatories such as HESS~\cite{HESS}, MAGIC~\cite{MAGIC} and VERITAS~\cite{VERITAS} consist of 2-5 telescopes. CTA will have an order of magnitude more telescopes, spread across two sites, one in the Northern and one in the Southern hemisphere.  The telescopes for CTA, which are currently in the final stages of development and prototyping, will be of three sizes:  Small-Size Telescopes (SSTs) with a diameter of 4 m which are sensitive in the highest energy ($>$ 10 TeV) range, Medium-Size Telescopes (MSTs) with diameters of 10-12m which will cover the core energy range (100 GeV to a few TeV), and Large-Size Telescopes (LSTs) of 23m in diameter, to detect the lowest energy photons starting at about 20 GeV. The small and medium size telescopes will each come in two designs, single mirror (Davies-Cotton type) and dual mirror (Schwarzschild-Couder).



This document describes the calibration methods and measurements of the single mirror MST prototype, in Adlershof, Berlin, carried out since 2012.

%
\section{CALIBRATION SETUP}
The MST prototype in Adlershof is equipped with 3 CCD cameras, 12 LEDs and numerous position and temperature sensors. Two of the CCD cameras are Prosilica GC1350s, with a chip size of 1360x1024 pixels  (4.65x4.65 $\mu$m$^2$ pixels) in individual Pelco housings. Of these, the \textit{SkyCCD} has an 85 mm Walimex Pro lens, giving a 4.26$^{\circ}$x 3.21$^{\circ}$ field of view (FoV), and is used to identify the true pointing direction of the telescope using astrometry. The \textit{LidCCD} with a 35 mm lens points along the optical axis and observes the focal plane and is used to determine the position of the Cherenkov camera using LEDs in the focal plane. The third CCD camera, called the \textit{SingleCCD} is of the type Apogee Aspen with a large FoV for simultaneous observations of the Cherenkov camera and the night sky, this camera is described in~\cite{1CCD}.  The \textit{LidCCD} and \textit{SingleCCD} cameras are also used for measurements of structural deformations of the telescope frame and optical support structure.

\section{CALIBRATION METHODS}
The following subsections describe the calibration of the MST prototype pointing and individual components of the telescope. 

\subsection{Pointing Calibration}

\begin{figure}[!htb]
  \centerline{\includegraphics[width=15cm]{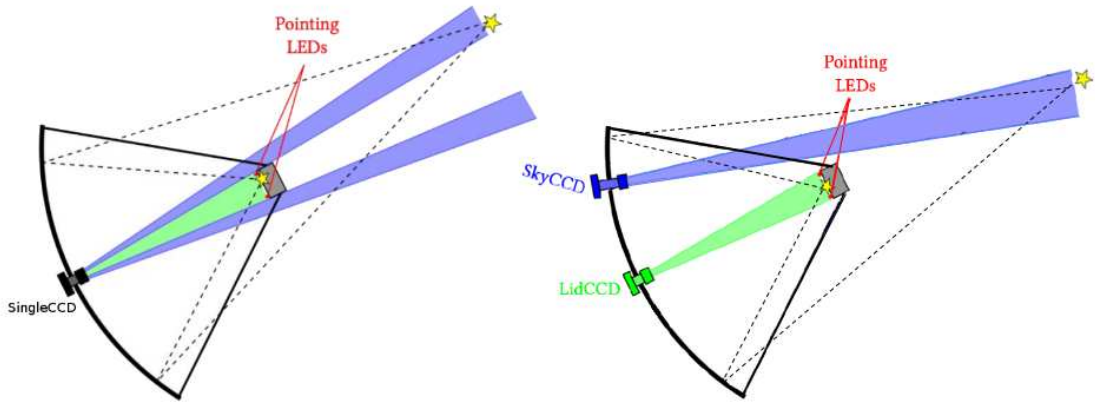}}
  \caption{The CCD camera layout for two pointing methods under study for the single mirror MSTs. Left - single CCD method, Right, Two-CCD camera method. Diagram adapted from~\cite{ccd1}.\label{fig1}}
\end{figure}

Figure \ref{fig1} shows the CCD camera layout for two alternative pointing calibration methods under consideration for the single mirror MSTs. The single CCD method, using one CCD camera located in the centre of the optical support structure, to observe the sky and the Cherenkov Camera simultaneously, is described in detail in~\cite{1CCD} and will not be discussed further here.  For the Two-CCD camera pointing method (shown on the right hand side of Fig.~\ref{fig1}), images from the \textit{SkyCCD} are analysed using Astrometry.net~\cite{astrom} software to identify the true pointing direction of this camera by matching the observed stars to indexed catalogs. Using the \textit{LidCCD}, the position of a reflected spot from a bright star on the Cherenkov camera housing can be measured with respect to the calibration LEDs, and thus a transformation between the \textit{SkyCCD} and \textit{LidCCD} FoV can be calculated. The true pointing direction of the optical axis of the telescope can then be calculated from the \textit{SkyCCD} images. A similar approach is already in use for precision pointing at the HESS experiment~\cite{hesspp}.

\begin{figure}[!htb]
  \centerline{\includegraphics[width=15cm]{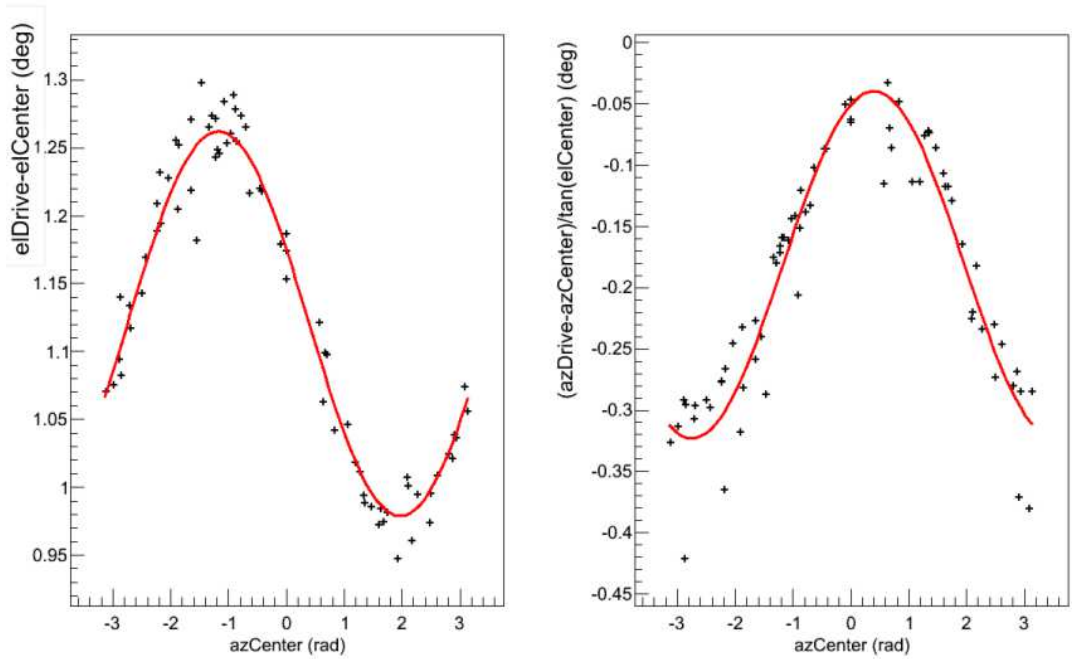}}
  \caption{Fits to altitude (left) and azimuth (right) dependence of deviations between the true and intended pointing directions in SkyCCD data with a 5-parameter bending model. \label{fig2}}
\end{figure}

In order to obtain a pointing model for the MST prototype, a preliminary bending model using the \textit{SkyCCD} has been developed using 70 \textit{SkyCCD} images. The difference between the intended pointing coordinates given to the telescope drive system, and the true pointing position as measured by Astrometry.net is fitted with respect to altitude and azimuth using a simple 5 parameter model. Figure~\ref{fig2} shows the results of these fits to the data, the parameters of which have been used to calibrate the drive system pointing direction. Since these results were calculated, significantly more pointing data has been taken in a campaign of nightly observations over several weeks; an updated bending model and calculation of the coordinate transformation to the telescope optical axis are in progress.  

\subsection{CCD Camera Calibration}
Prior to mounting on the MST prototype, the CCD cameras were assessed for any optical deformations such as chip movements or expansion during temperature changes.  Such distortion in the CCD cameras could lead to systematic uncertainties in the pointing calibration.  Temperature dependence studies were carried out in the climate chamber at DESY, Zeuthen.  

\begin{figure}[!htb]
  \centerline{\includegraphics[width=8cm]{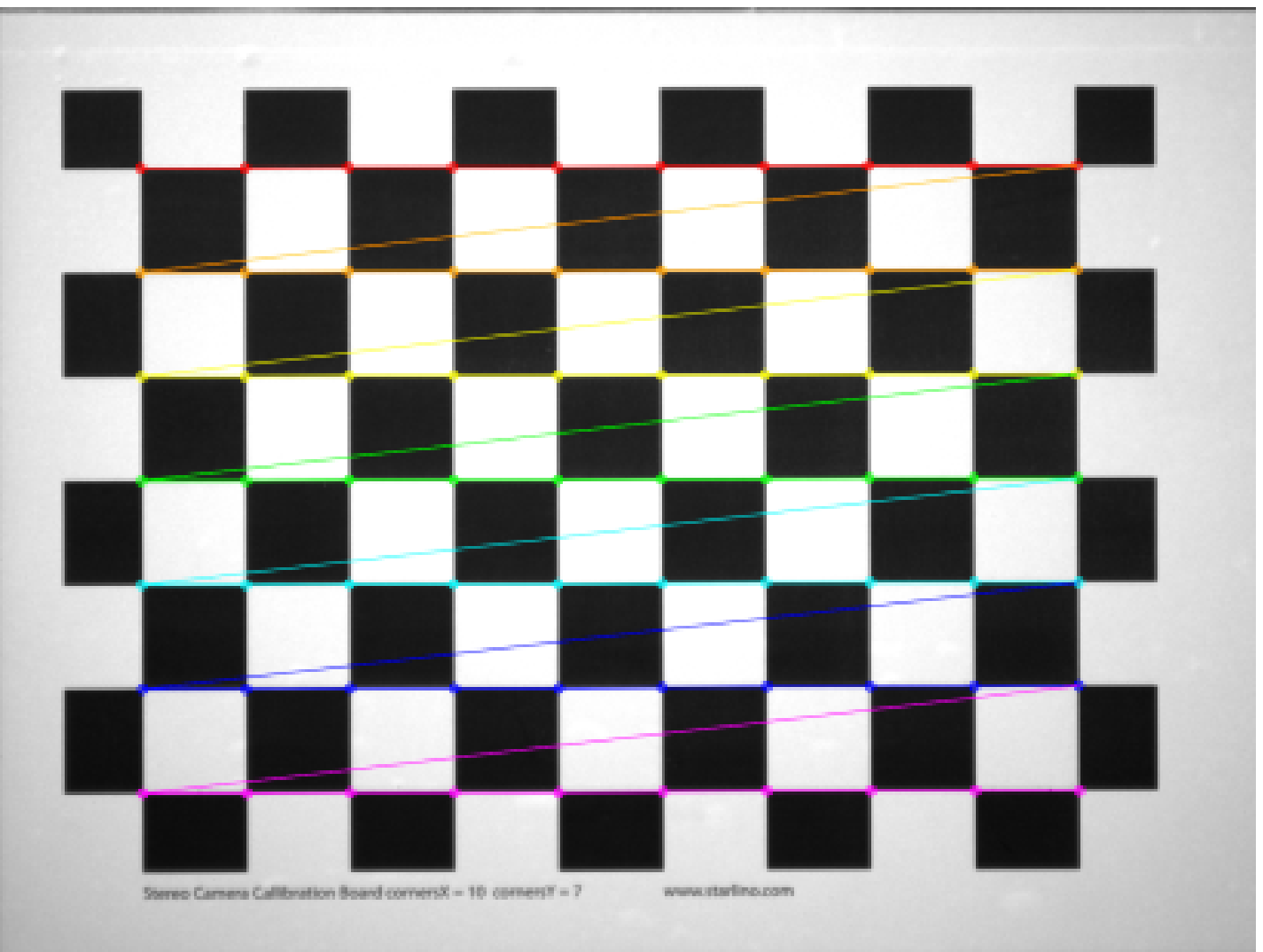}\includegraphics[width=8cm]{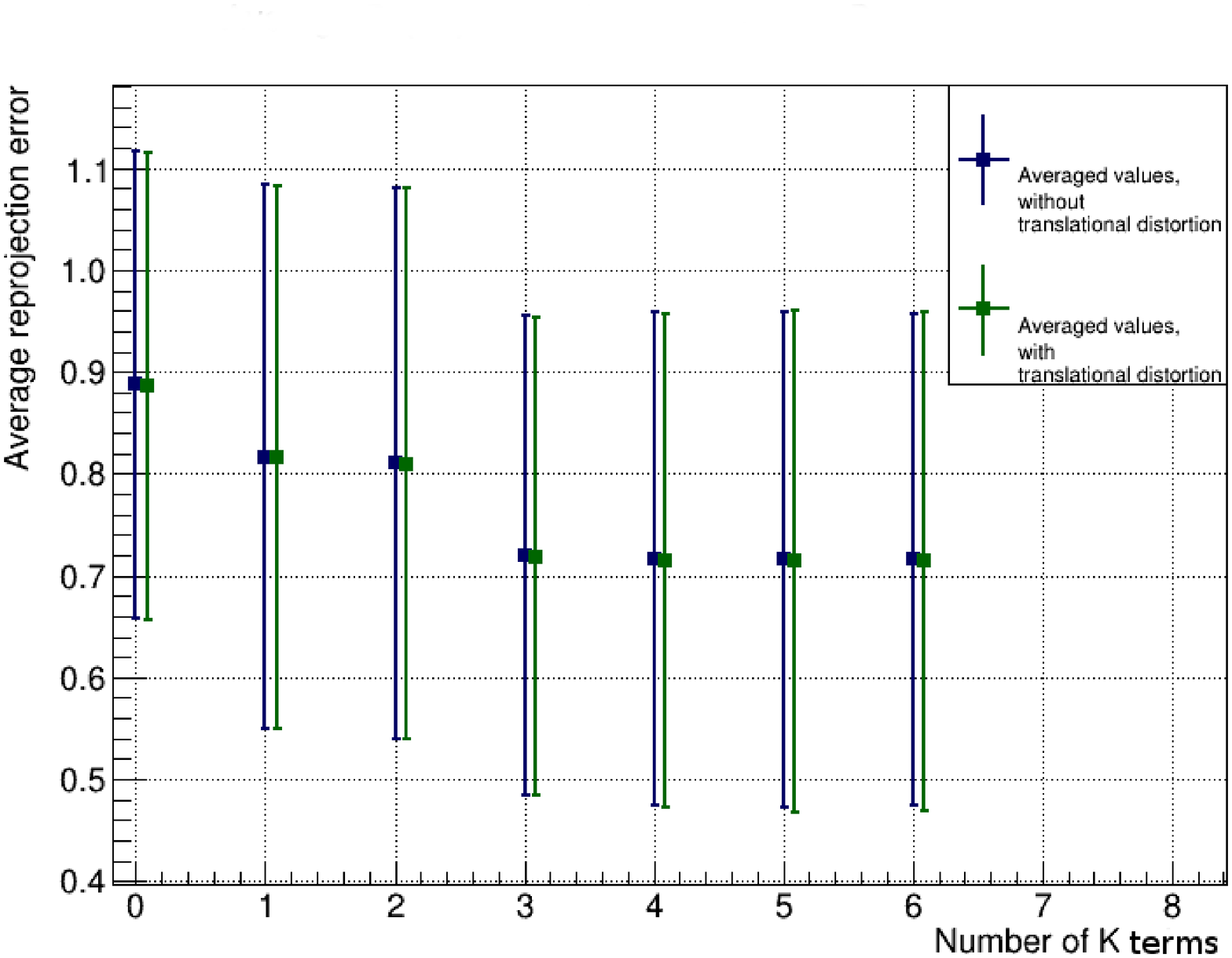}}
  \caption{Left - a chequerboard image taken with one of the CCD cameras under study,  OpenCV libraries were used to identify the corner points. Right - comparison of the goodness of fit for correction models with increasing numbers of radial distortion parameters. \label{fig3}}
\end{figure}

Image aberrations which would significantly affect the pointing calibration are those which cause an object to be shifted from its true position in the image plane. The optics (camera-lens system) were analysed and calibrated by fitting images using OpenCV libraries~\cite{OpenCV}. Radial and tangential deformations were fitted with several models with varying numbers of parameters, by analysing the corner positions of a chequerboard pattern imaged with the cameras. For each fit tens of images are taken with the chequerboard in a different position in the camera FoV.  An example of such an image is shown in Fig.~\ref{fig3}.  Also shown in Fig.~\ref{fig3} is the average reprojection error, a measure of the goodness of fit for a model used to calibrate the images, for each model~\cite{marcel}. 

\subsection{Mirror Alignment, the Bokeh Method}

The reflector for the MST is made up of multiple mirror segments. These segments must be well aligned to focus reflected light onto the Cherenkov camera and achieve the best possible Point Spread Function (PSF). To pre-align the mirror segments, a novel method called Bokeh alignment~\cite{bokeh}, pioneered by the FACT experiment, is used. This method has the advantage that it does not require a bright star or clear weather, and is possible to use at dusk or dawn as well as in darkness. 

\begin{figure}[!htb]
  \centerline{\includegraphics[width=15cm]{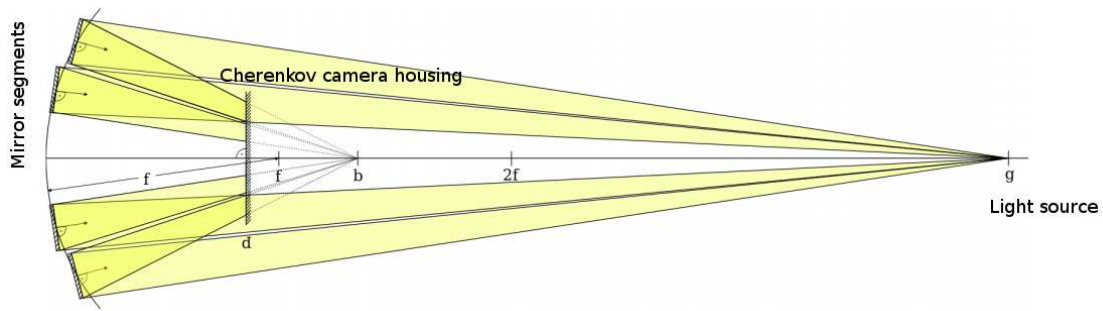}}
  \caption{The principle of Bokeh alignment. A point like light source is mounted at position g and reflected out of focus at point d. Diagram adapted from~\cite{bokeh}.   \label{fig4}}
\end{figure}

A bright light is mounted beyond the focal point of the optical structure and pointed at the reflector. This produces a blurred, out of focus, reflection from each mirror segment, known from photography as a Bokeh effect. Figure~\ref{fig4} shows a diagram of the set up required to achieve this effect. More traditional methods of mirror alignment require a very bright point source such as a star on a clear night. Conversely, with the Bokeh method, the reflected spots are easily located even in bad conditions. 

\begin{figure}[!htb]
  \centerline{\includegraphics[width=8cm]{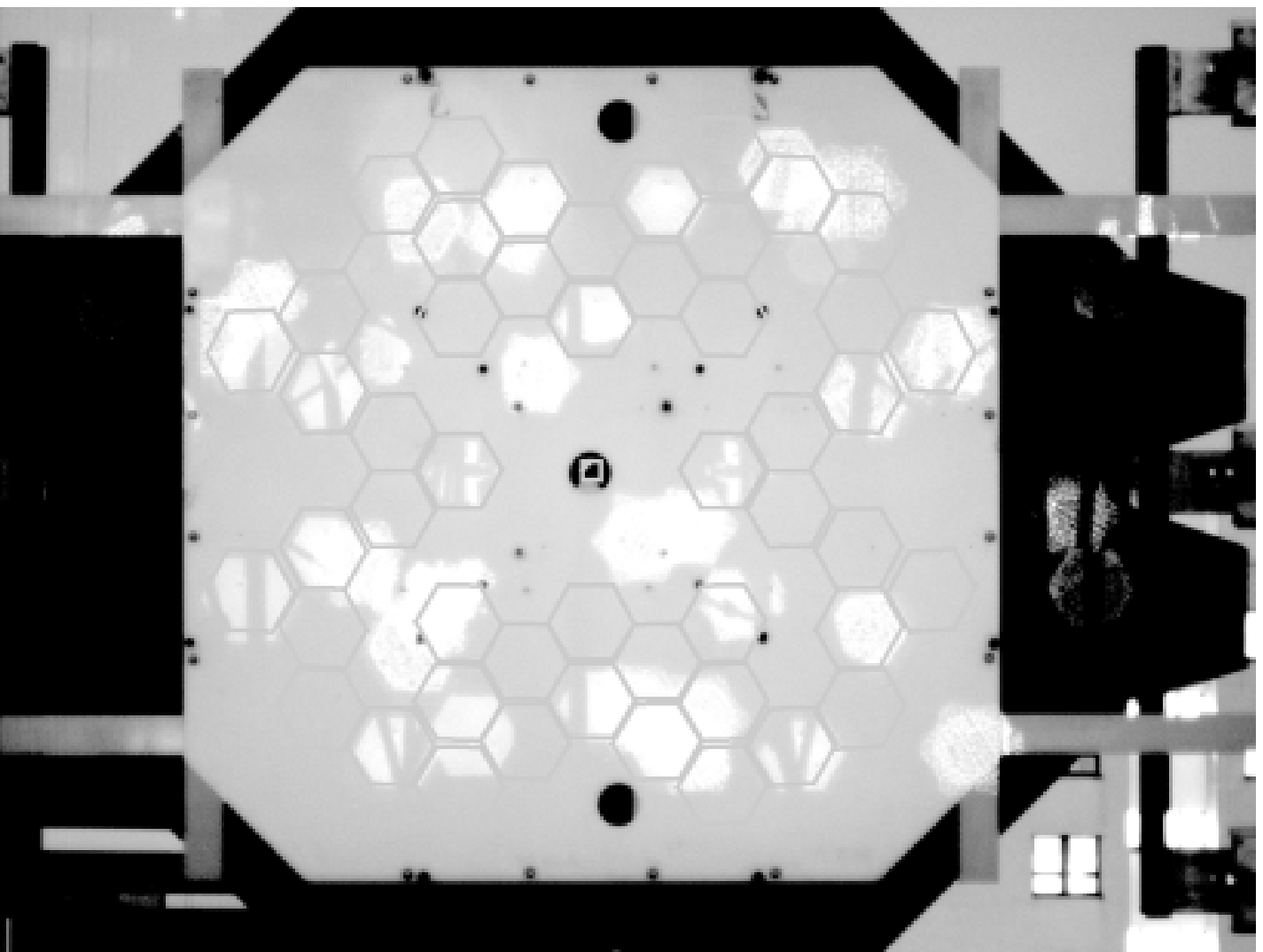}\includegraphics[width=8cm]{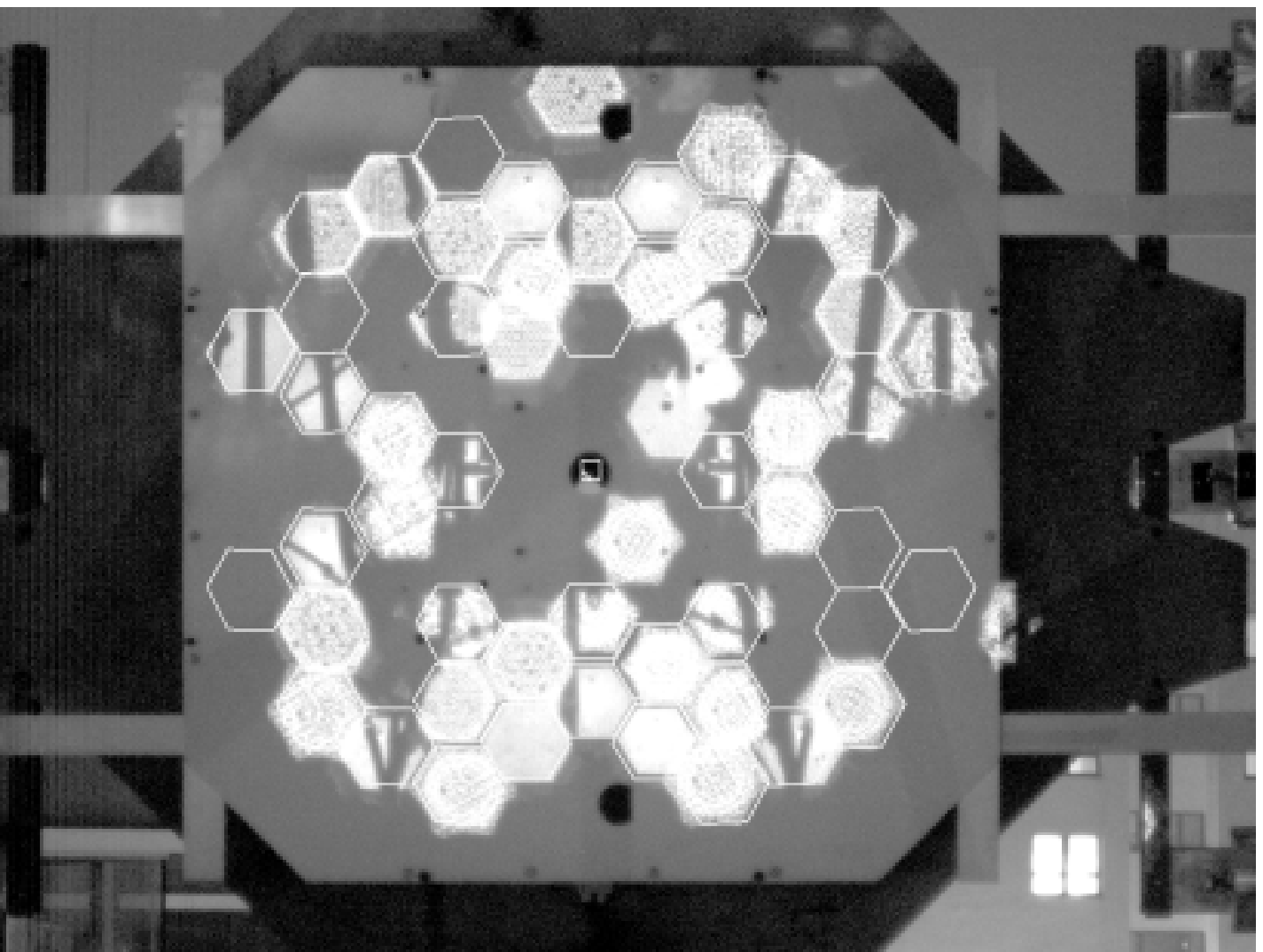}}
  \caption{Reflected Bokeh spots from mirror segments before (left) and after (right) alignment.  \label{fig5}}
\end{figure}

A template for the ideal positions of the Bokeh reflections from each segment when the mirrors are perfectly aligned can be produced using ray-tracing. For the MST prototype, this template was overlaid directly into the FoV of the \textit{LidCCD} in the CCD readout software (part of the ACTL package~\cite{actl}).  Each mirror segment can be adjusted in position by two actuators. By moving these actuators for each segment individually, the reflected Bokeh spots can be brought into the ideal position to match the template.  Figure~\ref{fig5} shows the reflected Bokeh spots before and after alignment with the template overlaid, as seen by the \textit{LidCCD}. The few spots which are still out of alignment on the ``after'' image are from mirror segments with non-functioning actuators. 

Bokeh alignment has been used successfully to bring the MST mirror segments into a good state of pre-alignment, which can later be improved using a point source reflection.

\section{SUMMARY}
A prototype MST for the CTA observatory has been constructed in Adlershof, Berlin, and operated for nearly 4 years. Individual components of the telescope instrumentation have been calibrated and tested.  The prototype has been used to develop and improve pointing models for single mirror MSTs, as well as for testing the Bokeh method of mirror alignment. These methods are being finalised in the end stage of prototyping and optimised to be applied at the MSTs during the commissioning phase of CTA.



\section{ACKNOWLEDGMENTS}
We gratefully acknowledge support from the agencies and organisations under Funding Agencies at www.cta-observatory.org.


\nocite{*}
\bibliographystyle{aipnum-cp}%
\bibliography{GAMMA2016_Oakes}%

\end{document}